\documentclass[epj]{svjour}
\usepackage{amssymb}
\usepackage{graphicx}

\begin{document}

\title{The physics of custody}
\author{ Andr\'es Gomberoff\inst{1}, V\'{\i}ctor Mu\~noz \inst{2}, and Pierre Paul Romagnoli\inst{3}}

\institute{
\inst{1} Universidad Andres Bello, Departamento de Ciencias F\'{\i}sicas, Av. Rep\'ublica 252, Santiago, Chile \\
\inst{2} Departamento de F\'{\i}sica, Facultad de Ciencias, Universidad de Chile, Santiago, Chile \\
\inst{3}  Universidad Andres Bello, Departamento de Matem\'aticas, Av. Rep\'ublica 252, Santiago, Chile
}

\abstract{
Divorced individuals face complex situations when they have children with different ex-partners, or even more, when their new partners have children of their own. In such cases, and when kids spend every other weekend with each parent, a practical problem emerges: Is it possible to have such a custody arrangement that every couple has either all of the kids together or no kids at all? We show that in general, it is not possible, but that the number of couples that do can be maximized.  The problem turns out to be equivalent to finding the ground state of a spin glass system, which is known to be equivalent to what is called a weighted max-cut problem in graph theory, and hence it is NP-Complete.
}
\maketitle
\section{Introduction}

The use of techniques borrowed from mathematics and physics in
tackling problems of social sciences has a long history. One
particularly fruitful field of study has been the analysis of social
networks~\cite{Wasserman,Borgatti}. There, individuals are treated as
vertices of a complex network. There is an interesting family of such
problems, where the properties of a society are related to the
individual interactions of the individuals in the same way
thermodynamics is related to microscopic pairwise interactions through
statistical mechanics~\cite{Castellano}. This point of view has been
used in many recent applications: evolutionary games \cite{Szolnoki},
social contagion \cite{Dodds}, conflict resolution~\cite{Torok}, among
others. In some cases, one may study the dynamics of a network by
minimizing the analog of an energy functional, that is, finding the
ground state of a physical system. This technique has been used, for
instance, in the study of social balance~\cite{Heider,Cartwright},
where the energy functional turns out to be the Hamiltonian of an
Ising spin system. In these models, the links between individuals may be of two kinds, ``friends" or ``foes". This family of systems has been   extensively studied until very recently~\cite{Antal,Radicchi,Radicchi_a,Marvel,Facchetti}.  They are called signed networks, and the minimum of the Hamiltonian measures the degree of tension produced by rivalries.

In this article we explore a similar kind of network, for which we will also minimize a spin glass Hamiltonian. The energy to be minimized will measure the degree of unhappiness of the whole network, and the minimization will proceed by choosing the orientation of 1/2-spin units with some given long range interactions. The precise definition of "unhappiness'' will be given in Sections 2 and 3 below. The methods used are very similar to the ones used in balance. The original problem, however, is quite different, and its relation with spin glasses and social balance is far from apparent. It has to do with the conformity of a network of divorced people in relation with the custody arrangement they agreed for being with their children on weekends. These agreements may be a source of discontent, specially nowadays, when it is common to have kids with two or more partners (this phenomenon, called multiple-partner fertility in the scientific literature, generates many other issues\cite{Meyer}). In those cases, it is usual to have a custody arrangement such that each parent enjoys the presence of the kids every other weekend. However, several inconveniences usually emerge: (i) not all the siblings are together the same weekend (when the parent has children with various partners); (ii) the parent is engaged in a new relationship with someone who also has kids, but they cannot get  all the kids together on the same weekend (along with a romantic one every other weekend). This issue may impose frustration and unhappiness in a quite large proportion of the population.

Unfortunately, the complex network of ex-wives and ex-husbands
people are waving makes difficult to reach an arrangement  that will
make everybody happy. In this note, we explain why, along with the conditions that may guarantee a happy solution. Moreover,
in the cases where a perfect solution may not be given, we study the
optimal arrangement, that would maximize the happiness of the group. Of course, we will give a precise definition of happiness 
It turns out, quite surprisingly,  that this problem is equivalent to find the ground state of a particular spin glass system~\cite{Sherrington}.

\section{Mathematical setting}

In order to simplify the computations, we will make use of some assumptions that do not appreciably change our conclusions.
First, we will assume only women-men couples. We will see that this makes our model much easier. In practice we may also justify it, because the proportion
of children of divorced gay and lesbian couples is currently quite small, and therefore the assumption is a good approximation

The domain of our problem is all the people in a population satisfying at least one of the following requirements:
\begin{itemize}
\item  Has children with two or more ex-partners.
\item  Has children with one or more ex-partners, and currently  married with a new partner who has kids from other relation.
\end{itemize}
Note that the world ``marriage'' is an abuse of notation here. It  is only a convenient way of referring to a
couple living together and with the desire of having all their kids together every other weekend.
Also note that the people who has children with only one ex-partner and is not married will not be affected by the problems listed in the introduction. They should always enjoy the presence of their kids together, because they only have one person (their ex-partner) to negotiate with (we assume that she or he should also want all siblings together).  Therefore, we do not take them into account. Same for people with kids from only one ex-partner and married with somebody with no kids or just common kids.

\begin{figure}[h]
\begin{center}
  \includegraphics[width=5cm]{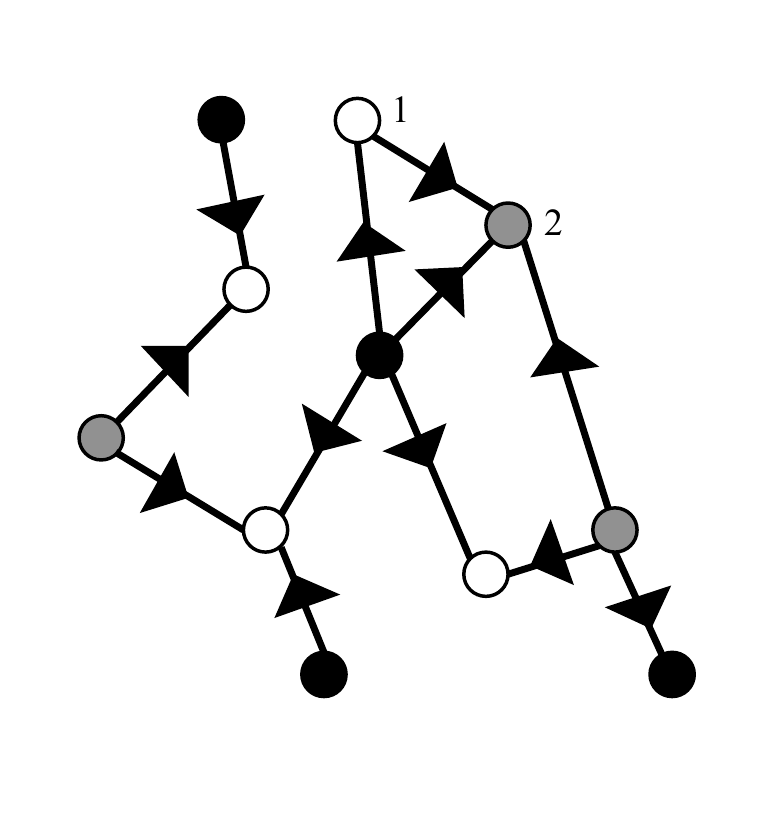}
  \caption{A custody arrangement state, CAS. We have labeled nodes 1 and 2 to clarify our definitions. The female 1 has kids with two men. One is single and has children with 4 women.  The other is married; the male of couple labeled 2. This CAS seems to be a source of discontent in female labeled 1, who has her kids visiting in different weekends.}
   \label{fig1}
   \end{center}
\end{figure}

We may now define a custody arrangement state (CAS) as an oriented graph, like the one depicted in Fig.~\ref{fig1}.  Black nodes are single males, white nodes are
single females and grey nodes represent married couples.  A connection between a black and a
white node means that both individuals have common children but are not a couple anymore.
Grey nodes may be connected to black or white ones. In the first case it means that the
female in the couple had kids with the individual represented by the
black node; in the second, the male in that couple had kids with the female
represented by the white node. Note that lines between grey nodes appear to be ill
defined. However, this is not important for the purpose of our study.
No matter who in the couple shares children with the corresponding
individual of the other couple, he (she) would want to have them all
together in the same weekend, and that is the only variable of
happiness in this study. (Note that grey nodes could be connected with two lines in case both members of one couple have kids with the corresponding member of the other.)
The orientation (arrows) in the edges of the graph points  towards the place the kids are going to spend any particular weekend. The CAS for the next weekend will have all the lines inverted (inverse orientation). An example is shown in Fig.~\ref{fig1}. We see that every individual and couple are happy with this CAS, except for female labeled 1. Her kids will not share weekends.

We would like to make a comment now about technical point. The man labeled 3 is not part of the domain of our problem as it was defined
at the beginning of this section. In fact, he is single and has kids with one ex-wife only. This means that he has will be happy with any arrangement. However, we keep it in the graph, because he is the reason his ex-wife is part of the domain. In that sense he may be understood as a "boundary" node of the graph.

Is it possible to find a CAS that brings happiness to every individual in the domain? The answer is not in general. Our goal is to find the best custody arrangement, that is, the CAS that minimizes the amount of unhappy people.

It is clear that the CAS of the whole planet must have disconnected
pieces, because not every pair of individuals is connected by a series
of ex-partnership links. We will therefore, without any loss of
generality, consider connected graphs only. Figure~\ref{fig1}
is an example of a connected CAS.

\section{Happy and convenient custody arrangement states }

Let us call a happy CAS one in which its orientations make everybody happy.
This means that at each node the arrows either all flow in or all flow out.
The conditions for existence of happy CAS are easily found. It is a well known fact of graph theory~\cite{Asratian} that these kind of graphs, called bipartite graphs, must have the following property: they may not have loops of odd length. Is easy to see why. Consider a happy CAS. The vertices in the graph may be divided in two groups: those which are only sources of arrows, and those which are only
sinks of arrows. If one starts a loop from a sink, the next step takes
us to a source, then to a sink, etc. When one closes the loop, back in
the starting sink,  it is clear that one must have made an even number
of steps.  The reciprocal is true as well, namely, any graph with no
loops of odd length is bipartite~\cite{Asratian}.
In general, however, the CAS will contain loops of  odd length, as we can see, for instance, in Fig.~\ref{fig1}, where the woman 1 and the couple 2 are part of a loop of length 3, which shows that there is no change of orientations that may transform this CAS into a happy one  (although we may transfer the unhappiness from woman 1 to any member of the loop). The problem people in this loop is facing is analog to the frustration fenomenon in an Ising antiferromagnet.
z
Note that if the graph is bipartite, that is, if there exists a set of orientations that make the CAS happy, then it is unique up to the reversal of all the arrows (which is the CAS happening the next weekend). This is obvious, because if the graph is connected, then after choosing the orientation of one edge, all the others would be immediately defined.

Unfortunately, the graph will not always be bipartite. However, there is a second best solution we may always construct. Not a happy CAS, but one we may call a ``convenient CAS". A convenient CAS is one in which
every individual enjoys the presence of all of her/his kids together. In this case, some couples may not be happy. Not all of their kids will enjoy every weekend together, but at least siblings will do.

To show that, note that in this case grey nodes are no longer
necessary. We cut the couples in their individual members (this means that, in contradistinction with the previous case, now we need to know exactly who are the parents in a line between two gray nodes). Each node is now an individual  female or male trying to have its arrows all flowing in or all flowing out. The procedure will, in general, leave a set of disconnected graphs. Now, however, the graphs are all bipartite, because the edges always connect black nodes with white nodes, so that any loop must be of even length. Therefore, there
exists a unique solution for each disconnected graph (up to reversal of all arrows).  Fig.~\ref{fig1} shows a solution for the example in Fig.~\ref{fig1}. Note that in this case, after cutting couples, there are two disconnected graphs left. There are, therefore, four different solutions that arise when inverting orientations of each disconnected piece. From those only two are really different solutions. The others will be the global change of orientation of these two.
\begin{figure}[h]
\begin{center}
  \includegraphics[width=5cm]{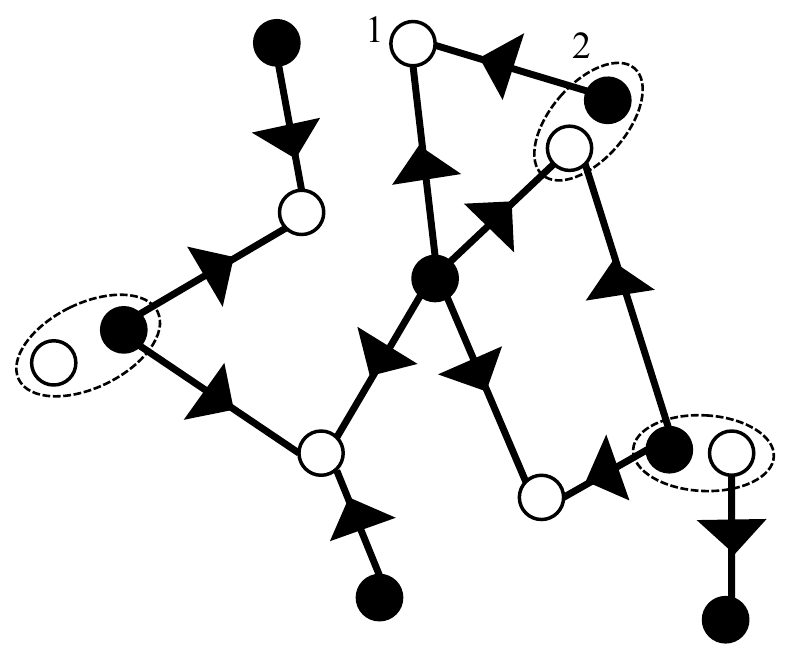}
  \caption{The grey nodes in Fig.~1 have been disentangled (the couples that were part of the grey nodes are grouped in dotted lines).  We end up with two disconnected graphs (we do not need to consider graphs of one individual). The restricted happy CAS has, therefore, 2 independent solutions.}
    \label{fig2}
   \end{center}
\end{figure}

In general, once we disentangle the grey nodes, the graph may end up in a set of  $N$ disconnected graphs, implying the existence of  $2^{N-1}$ different convenient CAS  solutions.

\section{Bringing the couples back}

We now bring the married couples back into the problem. Note that every
connected piece of our convenient CAS has every male or female on it
having the same kind of flow (either in or out). Let us call
``positively oriented" a connected subgraph in which the flow of arrows
in women is outwards, and ``negatively oriented" those in which women
have their arrow flow inwards. A key observation is that a marriage between individuals belonging to two graphs of the same orientation
will be always unhappy. This is because the woman in one graph and the
man in the other will have their arrows in opposite directions: they
will not have all of their kids together. As a direct consequence,
we see that couples belonging to the same connected piece will never
be happy. There is nothing we can do to avoid that.

\begin{figure}[h]
\begin{center}
  \includegraphics[width=4cm]{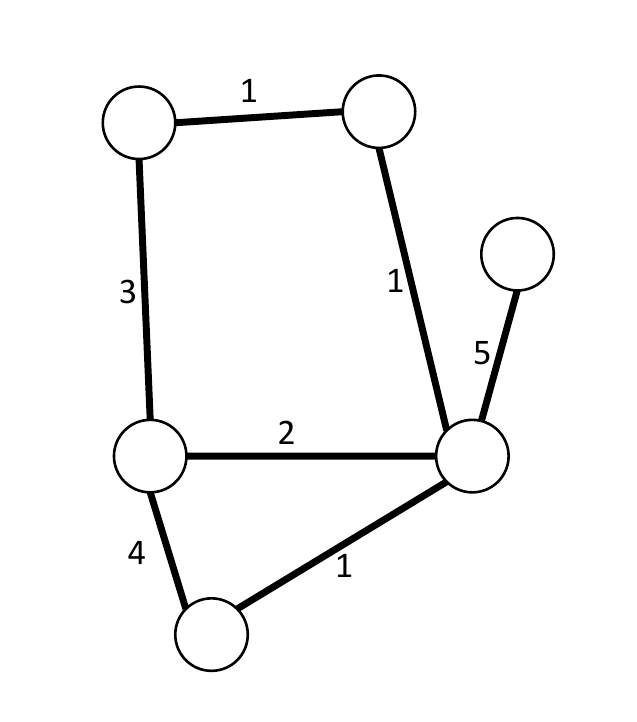}
  \caption{The graph of graphs. Each circle in this graph represents one connected subgraph
    in a convenient CAS. The orientations have been all removed. The lines between circles represent the presence of couples between the subgraphs. The weight of each line is the number of couples between these two subgraphs.}
 \label{fig3}
   \end{center}
\end{figure}

We may immediately read where the source of complete happiness
resides: every couple must be selected from disconnected pieces having
opposite orientations.  Having all this in mind we may reformulate the
problem of finding the optimal arrow orientation for a CAS. Start by
identifying all the disconnected pieces that emerge when married couples (grey
vertices) are disentangled in their two independent members. Let us
build a graph of graphs, in which each node represents each of the
connected subgraphs (see Fig.~\ref{fig3} for an example). Each
edge has a weight, representing the number of married couples with one member
from each of the corresponding vertices. The problem now is to choose
orientations for each subgraph, so that the maximum number of couples
is happy.  As before, if all loops have even length, then the graph of
graphs is bipartite, and we can choose the orientations to get a happy
CAS. If not, the problem  reduces to one which is well known in physics:  to find the ground state of a spin glass system. This problem, in turn, is equivalent to the so called weighted max-cut problem in graph theory \cite{Barahona}, which we will discuss in the next section.

Let us numerate the subgraphs $i=1,2,\ldots N$. We want to find the
happiest of the $2^{N-1}$ different convenient CAS  solutions. Call
$s_i$ the orientation of the $i$-th subgraph, which can only have two
values, say $+1$ or $-1$. The couples in a weighted edge are happy if
the orientations of their vertices are opposite, and unhappy if they
are the same. Call $J_{ij}$ the number of couples between node $i$
and $j$. We may then find a happiness function
\begin{equation} \label{H}
H=-\sum_{i>j}^{N} J_{ij}s_i s_j  \ .
\end{equation}
Since couples are given, weights $J_{ij}$ are fixed, whereas each CAS
corresponds to a certain configuration for $\{s_i\}$. Then, the
optimal solution can be found by optimizing $H$ with respect to the
set of configurations $\{s_i\}$.

Note that we have removed from the computation the case where both
couples are in the same node, because there is nothing we can do
with them. They are going to subtract a constant value of happiness to
every CAS in the family we are considering.  Also note what this
function is doing: It adds up the weights of all the edges between two
oppositely oriented vertices, and subtracts the rest. One may argue
that this is not quite the right function to maximize, for one should
simply count ``happy'' edges. Actually both ways are equivalent:
 If $F$ is the modified happiness function, where only ``happy" edges are taken into account in the sum, then it is straightforward to show that~\cite{Barahona}
 \begin{eqnarray}
F &=&-\sum_{i>j \ \mbox{\tiny opposite}}^{N} J_{ij}s_i s_j \nonumber \\
&=& \frac{1}{2}\left\{ - \sum_{i>j}^{N} J_{ij}s_i s_j  +\sum_{i>j}^{N} J_{ij}\right\}
=\frac{1}{2}H + C \ ,
\end{eqnarray}
where $C$ is a constant, having the same value for all CAS consistent
with the weighted graph. Therefore,
maximization of $F$ or $H$ inside this family of CAS is equivalent.

It is now clear that we need only change the sign of $H$ to see that
maximizing the happiness function is exactly equivalent to minimizing
the energy of a spin glass with $N$ Ising 1/2-spins, $s_i$, and long
range interactions $J_{ij}$ between them.

\section{Weighted max-cut problem and custody arrangements}

As discussed in the previous section, the equivalence between the custody arrangement problem and the spin glass ground state problem shows that it is in turn equivalent to the max-cut
problem,~\cite{Barahona} which is well known to mathematicians in graph theory.

The max-cut problem consists in finding a cut
of a given graph, that is, a continuous line that cuts it through it edges,
so that the sum of its weights is maximal. Thus, the cut divides the vertices
in two sets, which in our case correspond to the two orientations a
disconnected subgraph
may have. Therefore, the maximal cut maximizes
happiness, because the cut lines correspond to the ones connecting
vertices of opposite orientation, and therefore, to happy arrangements
for couples.

The problem is well known to be NP-complete~\cite{karp}, and there are
many algorithms and approximate methods to find
either local or global solutions, such as GC(max), Breakout Local Search
(BLS), MCFM~\cite{Ciesielski,Benlic,Wang_c,Lin,Ling}
However, it is interesting to notice that, in spite of the
computational complexity of the general problem, for some
graphs the max-cut problem can be polynomially solvable. It is the
case of planar graphs, that is, graphs where no edges intersect.
Fig.~\ref{fig3} shows a planar graph in fact, and thus this
particular problem should be solvable in polynomial time, and several
algorithms are available in that situation.~\cite{Shih,Berman,Liers}

\section{An example}

\begin{figure}[h]
\begin{center}
  \includegraphics[width=8.5cm]{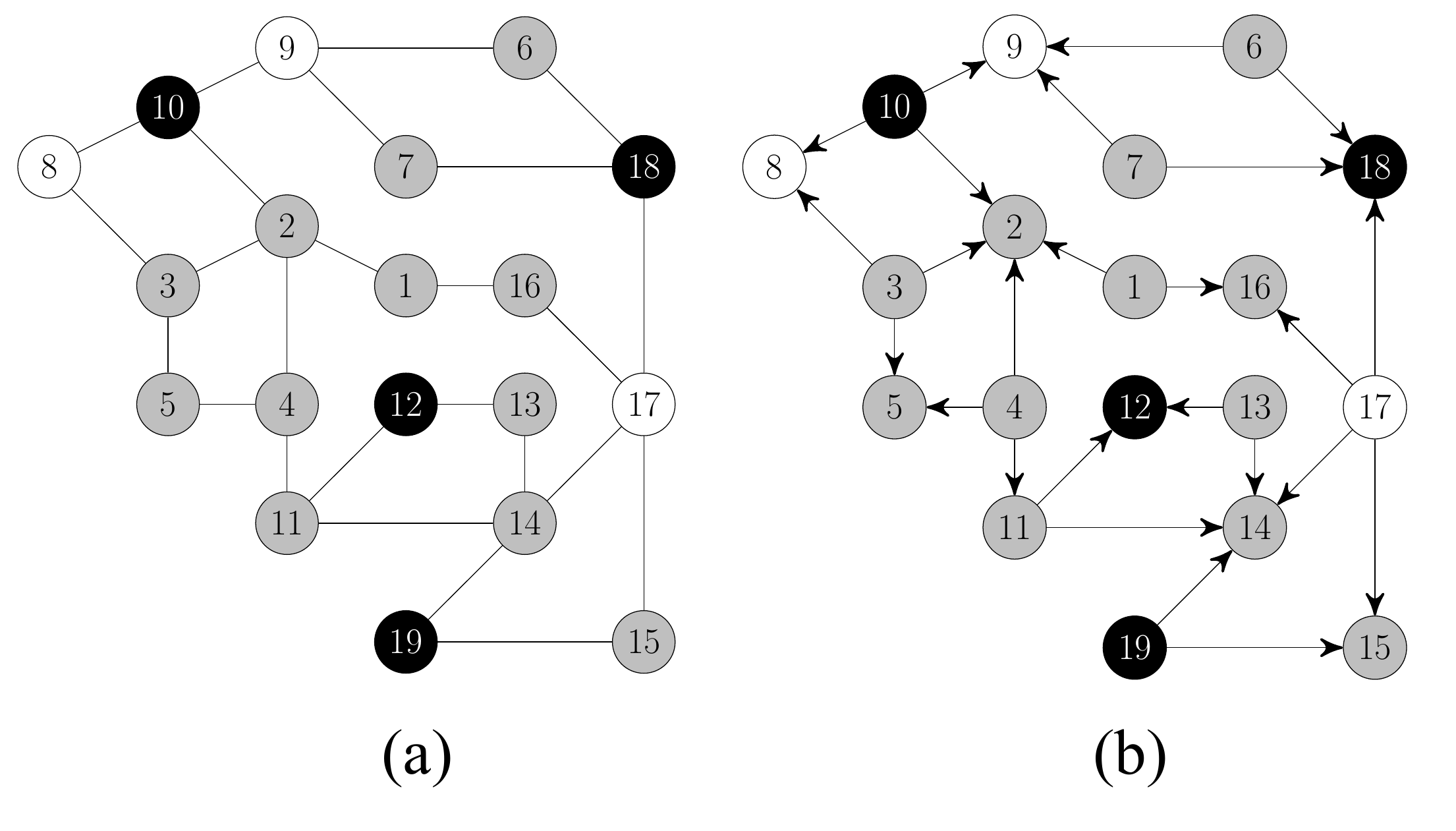}
\caption{A simple example of a graph depicting our problem. In (a) the problem is presented without orientations. In (b) we display the CAS that maximizes happiness. This solution is computed in the text using the methods described in the preceding sections. There is only one couple, namely 11, that cannot enjoy the presence of all children every other weekend.}
\label{fig4}
\end{center}
\end{figure}

To better understand the procedure, we now give a simple example. The graph in Fig. \ref{fig4}(a) is a posible network of ex-partners and couples. There are 19 nodes. Four of them represent single males (10,12,18, 19), three single females (8, 9,17) while the rest represent married couples.

\begin{figure}[h]
\begin{center}
  \includegraphics[width=8cm]{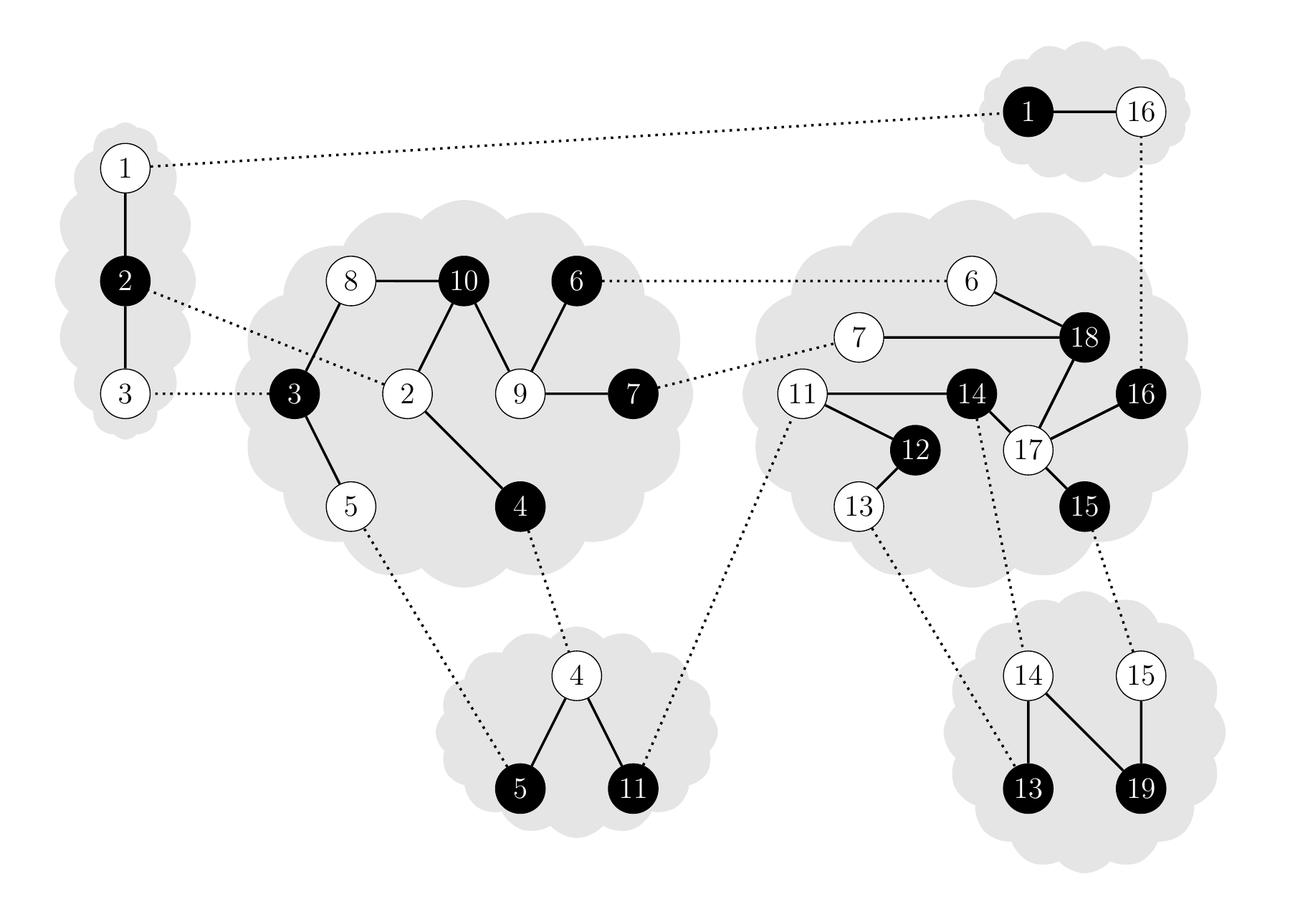}
  \caption{The couples (grey nodes) of Fig.\ref{fig4}(a) are broken in this graph. Both individuals are numbered with the same number and connected by a dotted line. The connected subgraphs are grouped in grey clouds. In this way, the graphs of graphs described in Sec.~4 emerges. It corresponds precisely to the one depicted in Fig.~\ref{fig3}.}
   \label{fig5}
   \end{center}
\end{figure}

There are several loops of odd lengths (for instance, 11-14-17-16-1-2-4), and therefore there cannot be a happy CAS solution. We therefore proceed by disentangling the couples in the search of a convenient CAS. For doing so, we need information not present in the graph of Fig.~\ref{fig4}, that is, the precise parents of children between connected gray nodes. An example of a disentangled graph giving rise to it is depicted in Fig. \ref{fig5}. The dotted lines connect couples, and the gray clouds group the connected pieces, which are precisely the nodes of the graph of graphs in Fig. \ref{fig3}.

\begin{figure}[h]
\begin{center}
  \includegraphics[width=8cm]{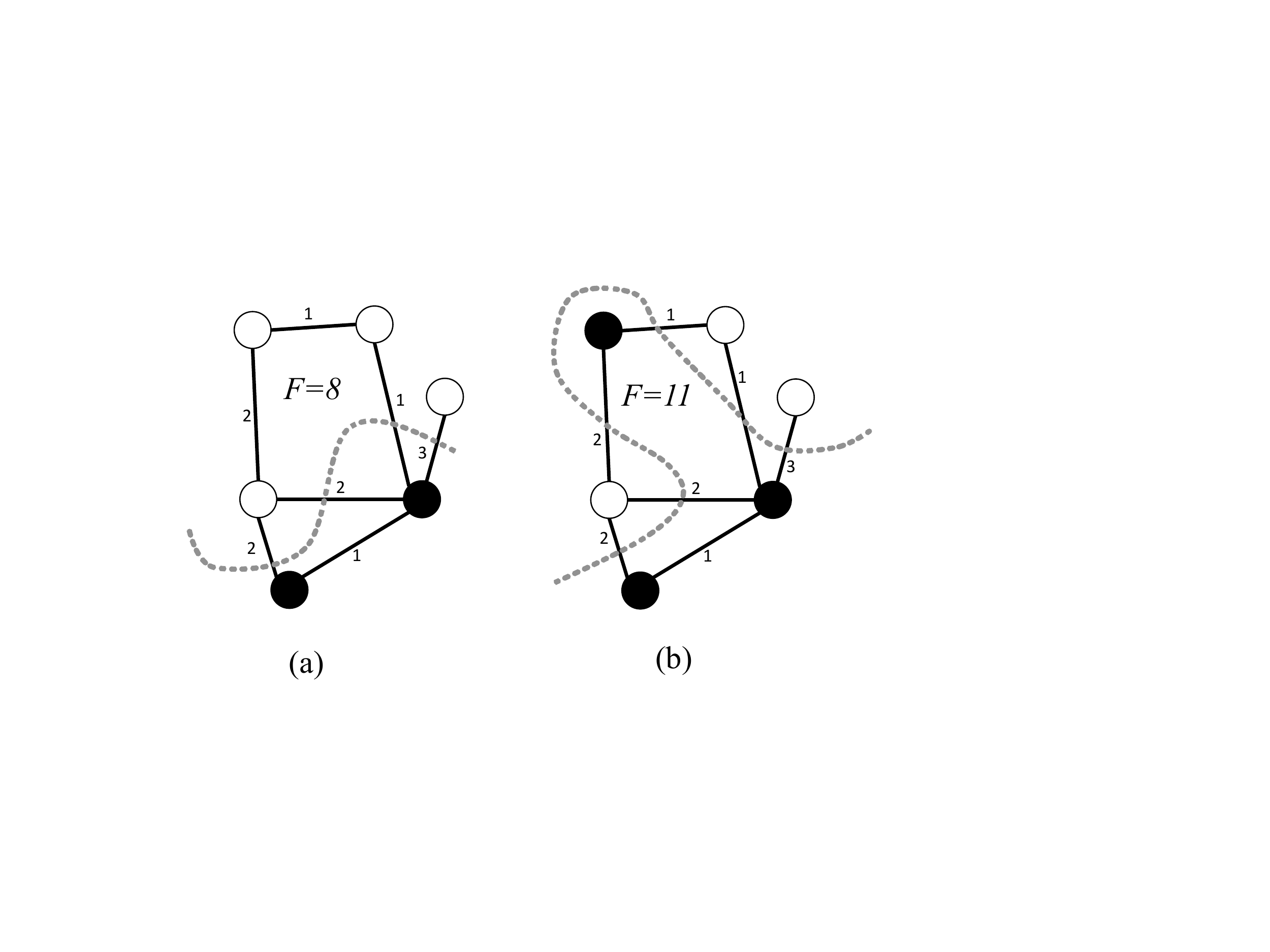}
  \caption{Two possible cuts (or orientation assignements) of the graph of graphs of Fig.~\ref{fig3} are shown. The happiness function $F$ is maximal for the second one, (b).  Black nodes represent positively oriented subgraphs, while white nodes represent negatively oriented subgraphs.}
    \label{fig6}
   \end{center}
\end{figure}

The problem then now reduces to find a max-cut solution of that weighted graph. In other words, we must find orientations of each node so that the happiness function (\ref{H}) is maximal. The 6 nodes of the graph imply $2^5=32$ different possibilities. In this simple case we may look at them all and find the solution by inspection. Two of these cuts are shown in Fig. \ref{fig6}. Positively oriented subgraphs are filled black, while negatively oriented are filled white. The second one of these graphs, with $F=11$ corresponds to the maximal solution.
This solution is the one displayed in Fig. \ref{fig4}(b).
One sees that only one couple is going to be unhappy with the arrangement. It is the one connecting two black subgraphs in \ref{fig4}(b), and labeled 11 in Fig. \ref{fig4}(b). We may corroborate however, by lloking at Fig. \ref{fig5}, that in this solution siblings will always be together, as they should.

\section{Summary}
\label{summary}

In this note we considered the problem of custody arrangements between divorced couples,
which very often state that children spend every other weekend with each parent.
A graph model for the configurations of custody arrangements for
divorced couples with children is presented. In the graph, nodes
represent married couples and individuals, and a link between two individuals
shows that there are kids in common. Links
are oriented in the direction of the parent enjoying the company of his kids on a given weekend.
The resulting oriented graph is called a CAS (Custody Arrangement State).
If an individual enjoys the presence of all of his kids together every other weekend we call him a happy individual.
The same for couples having all of their kids together. Hence, in the graphical representation, a happy node (males, females or married couples) is one
in which its edges either all flow in or all flow out of it.
A happy CAS is one in which all nodes are happy. One may choose orientations of the edges such that the CAS gets happy if every loop  in the graph has an even number of links. This is not always the case.
However, we have shown that even for unhappy CAS, a ``convenient'' state may be found,
where all individuals have their respective
children with them every other weekend, but some couples may not have
all their children together.

When happy CAS exists it is unique up to reversal of overall orientation. Convenient CAS, however, are not unique. One should choose between all of them for the one where the number of happy couples is maximized. We have shown that this is equivalent to the problem of finding the ground state of an Ising model for a
spin-glass. In turn, it is known that this amounts to solve a weighted max-cut problem in graph theory.

It would be interesting to estimate the ``unhappiness'',  of an actual society, as defined in
our paper, and apply the algorithm to it. 
To get an estimation of the size of the problem we take some statistics from the literature and feed
a simple simulation using them. The details of the simulation will be given elsewhere. According to \cite{Dorius}, the prevalence of multiple partner fertility in women in the United States is $22\%$
among mothers aged 41-49 (data from 2006). In that same paper it is found that $16\%$ of women in this age range have no children.  Now, according to \cite{Guzzo}, the prevalence of multiple partner fertitlity in fathers in the United States is $17\%$ (data from 2002).  
At this stage we are only interested in a coarse estimation, therefore we run a random graph simulation with a group of 10,000 individuals, assuming that $16\%$ of the women have no kids and that $20\%$ of both men an women have kids with more than one partner. We further assume that multiple partner fertility in women is always with two partners (most of them are, so this is a reasonable simplification. An artifact in the algorithm makes it possible for men to have kids with more than two partners with a small probability). We assume that $90\%$ of people are in a relation and that there is $85\%$ of chance that this relation is with one of the partners the individual has children with. We note that in these circumstances, most of the connected pieces of the graph have a small number of nodes (and all of them less than 10). The corresponding graph of graphs is disconnected and the probability of ending up having loops is quite small. This would mean that, for a typical group of people inside the U.S., a happy solution should be possible.
However, if one considers some specific groups, this is not necessarily true. When the chance of multiple partner fertility gets bigger than $30\%$, loops begin to become a common property of the graph of graphs. Actually, these numbers are not uncommon in some ethnic or socio-economic populations, as one may see, for instance, in \cite{Guzzo}.
Of course, we have done many simplifying assumptions, but our
preliminary numerical experiments suggest that unhappiness, as defined
here, may be a relevant problem if population is small (closed
communities, small towns, etc.) or if the prevalence of multiple parter fertility gets  bigger. Also, for big populations, we would expect that deviations will 
cause a considerable number of big networks. 
We plan to investigate in detail these issues in the future.

Of course, in real life, many obstructions not considered in this note, may emerge.
Let us see some examples. When gay and lesbian couples are included, even the
connected subgraphs may have closed loops with odd length. In this case, we must start by maximizing happiness on each subgraph. This may be done, again, by mapping it into a spin glass model. Now, all the edges will be of the same weight, and therefore the interaction between spins is either zero or one.
Another real life problem emerges when there are obstructions for some individuals on the weekend they may be with the kids (someone that, for instance, must work every other weekend). In the magnetic analog this represents a spin whose orientation is fixed. In that case, again, we will be forced to maximize happiness at each subgraph, with given constraints. This process may end up fixing the orientation of the entire subgraph.

Other difficulties include cases where, for instance, an individual
would not prioritize, as we do here, to have siblings together, and
would choose to have one of her/his kids with the kids of her/his spouse
instead. Also we may face the fact that these graphs are not static.
New couples are constantly forming, while others disappear and new kids
are born. Finally, one may wonder that, even if a happy solution
exists for our own custody arrangement network, it would be impossible
in practice to organize all the people involved.

We think that these problems are an important source of stress in modern life, and it is important and interesting to address them in the future.

\section{Acknowledgments}
We would like to thank Jos\'e Aliste, Natalia Mackenzie, Iv\'an Rapaport, Crist\'obal Rojas and Jos\'e Zamora  for valuable discussions.
VM thanks the financial support of Fondecyt under
grant No.~1121144.

\end{document}